**Specifics about Specific Ion Adsorption from Heterodyne-Detected Second Harmonic Generation**

Mavis D. Boamah[1], Paul E. Ohno[1], Emilie Lozier[1], Jacqueline Van Ardenne[2], and

Franz M. Geiger[1]*

[1]Department of Chemistry, Northwestern University, Evanston, IL 60208

[2]Department of Chemistry, University of Puget Sound, Tacoma, WA 98416, USA

*Corresponding author: geigerf@chem.northwestern.edu

**Abstract.** Ion specific outcomes at aqueous interfaces remain among the most enigmatic phenomena in interfacial chemistry. Here, charged fused silica/water interfaces have been probed by homodyne- and heterodyne-detected (HD) second harmonic generation (SHG) spectroscopy at pH 7 and pH 5.8 and for concentrations of LiCl, NaCl, NaBr, NaI, KCl, RbCl, and CsCl ranging from 10 μM to several 100 mM. For ionic strengths around 0.1 mM to 1 mM, SHG intensities increase reversibly by up to 15% compared to the condition of zero added salt because of optical phase matching and electrical double layer. For ionic strengths above 1 mM, use of any combination of cations and anions produces decreases in SHG response by as much as 50%, trending with ion softness when compared to the condition of zero added salt. Gouy-Chapman model fits to homodyned SHG intensities for the alkali halides studied here show charge densities increase significantly with decreasing cation size. HD-SHG measurements indicate diffuse layer properties probed by the SHG process are invariant with ion identity, while Stern layer properties, as reported by $\chi^{(2)}$, are subject to ion specificity for the ions surveyed in this work in the order of $\chi^{(2)}_{RbCl} = \frac{1}{2}\chi^{(2)}_{NaCl} = \frac{1}{4}\chi^{(2)}_{NaI}$.

*Corresponding author: geigerf@chem.northwestern.edu

**I. Introduction.** While ion-specific interactions at charged interfaces have important consequences for adsorption, protonation/deprotonation, interfacial charge densities and potentials, and the structuring of interfacial species within the electrical double layer, they remain one of the most enigmatic phenomena to describe with microscopic models. Part of the issue is rooted in the fact that descriptions of the electrical double layer remain largely confined to Bragg-Williams (mean field) approximations in the form of 'layer cake' and 'primitive ion' models.[1-3] Experimental approaches that provide microscopic properties comparable to what is possible through computer simulations of aqueous interfaces,[4-11] which could lead to an increased molecular understanding, control, and prediction of interfacial phenomena over charged interfaces, remain elusive, especially for ionic strengths between 10 μM - 0.1 M. As a result, it is challenging to interpret experiments probing differences in, for instance, how NaCl vs CsCl interact with a charged aqueous interface. Compounding the complexity of the problem is the observation from electrokinetic measurements that the interfacial charge density of certain oxides varies with ionic strength, at least for the ionic strengths accessible in those experiments.[42-44,117]

Despite these challenges, recent years have seen some significant progress: For instance, ion identity (element and charge), electrolyte concentration, and bulk solution pH are some of the factors that are now known to control surface charge density and the structure of interfacial water molecules and the hydrogen-bond networks they establish within the electrical double layer.[12-17] Moreover, potentiometric measurements and calculations with surface complexation models have established that cations and anions influence charge densities at silica/water interfaces via their adsorption affinities.[18-20] Likewise, nonlinear optical measurements have revealed that halide and alkali ions influence silanol group dissociation at silica/water interfaces,[21-23] a result that perhaps is linked to reports that certain alkali chlorides at high ionic strengths enhance silica dissolution rates.[18, 24-28] Perturbations to the hydrogen-bonded network of interfacial water molecules[29] and a considerable lengthening of the $T_2$ relaxation time of interfacial water molecules with certain ions at elevated ionic strength[30-32] have been reported as well. We now investigate how a series of alkali halides interact with the fused silica/water interface at concentrations ranging from 10 μM to 0.1 M at circumneutral pH. Specifically, we investigate ion-specific outcomes in second harmonic generation (SHG) spectroscopy[33-35] featuring, for the first time, homodyne and heterodyne-detection (HD) using the Eisenthal $\chi^{(3)}$ method under off-resonant conditions[36-41] using a newly developed HD-SHG spectrometer.[42] When the frequency of the incident fundamental electric field, $E_\omega$, is tuned off electronic or vibrational resonances, the second harmonic electric field, $E_{2\omega}$, produced at the interface is described by the following model:[7, 13, 42-53]

$$E_{2\omega} \propto \chi^{(2)} E_\omega E_\omega + \chi^{(3)} E_\omega E_\omega \int_0^\infty E_{dc}(z) e^{i\Delta k_z z} dz \qquad (1)$$

Under off-resonant conditions, the second- and third-order nonlinear susceptibilities of the interface, $\chi^{(2)}$ and $\chi^{(3)}$, respectively, are purely real. The factor $\Delta k_z$ is the wave vector mismatch describing the inverse of the coherence length of the SHG process (1.1 x 10$^7$ m$^{-1}$ in our case),[42] and $E_{dc}$ is the $z$-(depth) dependent electric field produced by interfacial charges and ionic screening, given by $E_{dc} = -d\Phi(z)/dz$, where $\Phi(z)$ is the electrostatic potential.[49-52] This dependence of the SHG response on the electrostatic potential has led to its use in the form of an "optical voltmeter" for probing the electrical double layer at insulator/aqueous interfaces.[39, 54]

If $\Phi(z=0)$ is expressed as the Gouy-Chapman model (explained further in the Supporting Information), the homodyne SHG response corrected for phase matching is given as follows:

$$\left|(E_{2\omega} - A)/\frac{\kappa}{\kappa - i\Delta k_z}\right|^2 = \left|B\left\{0.05139\, V\, \text{arcsinh}\left(\frac{\sigma[Cm^{-2}]}{0.1174\left[C\, m^{-2}\, M^{-\frac{1}{2}}\right]\sqrt{C[M]}}\right)\right\}\right|^2 \quad (2)$$

Here, A = $\chi^{(2)}E_\omega E_\omega$ with units of V m$^{-1}$, B = $\chi^{(3)}E_\omega E_\omega$ with units of m$^{-1}$, and σ is the surface charge density reported in Cm$^{-2}$. We normalize all E$_{SHG}$ data to the SHG response recorded at high ionic strength (where charge screening and, therefore, the surface potential is minimal). Thus, for our data, the value of A is effectively unitless whereas B has the units of V$^{-1}$. Using eqn. (2), we recently reported that "models recapitulating the experimental observations are the ones in which (1) the relative permittivity of the diffuse layer is that of bulk water, with other possible values as low as 30, (2) the surface charge density varies with salt concentration, and (3) the charge in the Stern layer or its thickness varies with salt concentration."[13]

We now investigate ion specificity in the observed SHG responses by varying the nature of the alkali ions from Na$^+$ to K$^+$ to Rb$^+$ and Cs$^+$ and that of the halides from Cl$^-$ to Br$^-$ and I$^-$. We work at pH 7, where the fused silica/water interface is slightly negatively charged ($\sigma_O$=0.015 C m$^{-2}$).[36, 55-56] We correct for phase matching, which is relevant for reflection geometries zat low ionic strength conditions[13, 49-52, 57] by determining the square modulus value of the SHG E-field minus the average value of the SHG E-field at high ionic strengths (≥ 0.1M) divided by $\frac{\kappa}{\kappa - i\Delta k_z}$ as described and published in the SI of our prior work.[13] We then conduct phase-referenced ("heterodyne-detected") HD-SHG measurements[42] to separate the $\chi^{(2)}$ and $\chi^{(3)}\Phi_0$ contributions for NaCl, RbCl, and NaI at constant low ionic strength (10 μM) so as to analyze the ion-specific nonlinear optical responses without having to rely on Gouy-Chapman theory to express the interfacial potential. To this end, Equation 1 is rewritten as follows:[13, 42, 49-52, 57]

$$E_{2\omega} \propto \chi^{(2)}E_\omega E_\omega + \chi^{(3)}E_\omega E_\omega \Phi(0)\cos(\varphi_{DC})e^{i\varphi_{DC}} \quad (3)$$

For a DC-field falling off exponentially over the Debye screening length, $\lambda_D$, the phase angle ($\varphi_{dc}$) of the $\chi^{(3)}$ term takes the form[42]

$$\varphi_{DC} = \arctan[\Delta k_z \lambda_D] \qquad (4)$$

Here, $\varphi_{DC}$ is the phase angle associated with the z-dependence of the $\chi^{(3)}$ signal (i.e., produced at varying depths away from the interface) and $\Delta k_z$ is again the wave-vector mismatch, which is given by the wavelengths of the input and output wavelengths and the relevant refractive indices (again, $\Delta k_z = 1.1 \cdot 10^7$ m$^{-1}$ for our experimental conditions).[42] To obtain $\chi^{(2)}$ and $\chi^{(3)}$, we compute $\varphi_{DC}$ = 13.3° at 0.2 mM ionic strength using Debye Hückel theory for a 1:1 electrolyte to describe the Debye length in eqn. 4. We then reference the phase of the SHG response detected at the PMT ("$\varphi_{sig}$ of $E_{sig}$") using an external piece of quartz as the local oscillator. As illustrated in our recent publication on HD-SHG,[42] equations 3 and 4 can be rewritten as follows:

$$\chi^{(2)} = \cos(\varphi_{sig}) E_{sig} - \cos^2(\varphi_{DC}) \chi^{(3)} \Phi_0 \qquad (5)$$

$$\chi^{(3)} \Phi_0 = \frac{\sin(\varphi_{sig}) E_{sig}}{\cos(\varphi_{DC}) \sin(\varphi_{DC})} \qquad (6)$$

The $\chi^{(2)}$ contributio is due to the populations and nonlinear hyperpolarizabilities of interfacial species, including the water molecules, the surface silanol groups, and the adsorbed ions. By means of comparison with the triple layer, or Gouy-Chapman-Stern model,[55, 58-61] $\chi^{(2)}$ is currently understood[7-8, 16, 51-53] to describe the physics and chemistry in the Stern layer. The $\chi^{(3)}$ term, in turn, has two origins: (1) the third-order hyperpolarizability of water molecules, $\alpha^{(3)}$, and (2) the polarization of water molecules by the interfacial potential, $\Phi_0$,[37, 62-63] in the diffuse layer. The expression for $\chi^{(3)}$ is given as:[62]

$$\chi^{(3)} = n_{H_2O}\left(\alpha^{(3)}_{H_2O} + \mu \alpha^{(2)}_{H_2O}(bkT)^{-1}\right) \qquad (7)$$

Here, $n_{H_2O}$ is the surface density of water, μ is the permanent dipole moment of water, b is a constant determined by the susceptibility element, and $\alpha^{(2)}$ is the second order molecular hyperpolarizability. The permanent dipole moment of water results from how the different dipoles of the water molecules present at the interface and bulk add up vector-wise.

## II. Experimental Methods.

**A. Flow Set-Up for Homodyne SHG Measurements.** For homodyne SHG measurements, we clamped a UV-grade fused silica hemisphere (ISP Optics, 2.54 cm diameter, QU-HS-25-1) leak tight onto a Teflon flow cell having an internal sample volume of 2 mL using a Viton O-ring as described, in detail, elsewhere.[64] We use peristaltic pumps configured to maintain conditions of creeping flow and low shear rates at constant flow rates of ~1 mL/sec within the several mL internal volume of the sample flow cell.

**B. Flow Set-Up for Phase-Referenced SHG Measurements.** For heterodyne measurements, a UV-grade fused silica hemisphere (ISP Optics, 2.54 cm diameter, QU-HS-25-1) is clamped leak tight onto a flow cell with a PTFE inner wetted area using a fluoroelastomer O-ring.[42] We use peristaltic pumps configured to maintain conditions of creeping flow and low shear rates at constant flow rates of ~5 mL/min across the internal volume of the sample flow cell. We cycle at least three times between 0.2mM of each salt and $CO_2$ equilibrated water at pH 5.8 and recorded interference patterns. In this manner, a minimum of three interference scans lasting for a total of 15 minutes (5 min per individual scan) are recorded for each water flow or salt flow in each cycle. Measurements are conducted for the different salts (NaCl, RbCl, and NaI) on freshly prepared hemispheres on different days.

**C. Homodyne SHG Measurements.** We reported earlier the experimental procedure for our homodyne SHG studies at the fused silica/water.[50] Briefly, we probe the interface at a 60° incident angle from the surface normal using 120 femtosecond pulses of an 82 MHz Ti:Sapphire oscillator at 800 nm (Mai Tai, Spectra-Physics). We maintain the input laser power measured before the sample stage at 0.46 ± 0.05 W. We set the polarization combination to *s*-in/all-out in all cases except where stated otherwise. A single photon counter (SR400, Stanford Research Systems) detected the SHG signal, after which the signal was averaged using a boxcar (sliding-average) procedure written in IgorPro (Wavemetrics). At least three replicate adsorption isotherms were collected individually for each alkali halide on multiple days and using various fused silica substrates at laboratory temperature (21°C - 22°C).

**D. Heterodyne-Detected SHG Measurements (HD-SHG).** We recently detailed the set-up for our phase-referenced, or heterodyne-detected, HD-SHG experiments.[42] The measurements involve recording interference patterns, which are the result of coherent interference between the SHG signal generated from the sample and a local oscillator (LO). The LO is generated within a 50 μm z-cut α-quartz crystal. The measured signal is due to the translation of the reference quartz crystal along the beam path of the SHG response (from the sample stage), to give an interference pattern. Hence, the interference patterns or scans are SHG intensities recorded as a function of stage position.

In the experiments, we probe the fused silica/water interface using 200 fs laser beam pulses produced at a repetition rate of 200 kHz at 1030 nm. The pulse energy is kept below 0.4 μJ (80 mW). We maintain the polarization combination at s-in/p-out for all phase-referenced measurements and use a Hamamatsu PMT combined with a preamplifier and gated photon counter (SRS, SR400) employing a gate width of 50 ns for detection. Using the second channel of the SR400 with the gate offset from the laser pulses, we subtract spurious (dark counts) counts from the detected signal intensity on the fly.

**E. Alkali Halide Solutions.** Dilute solutions of NaOH and HCl are used to maintain pH 7 of the alkali halide solutions except for CsCl, where CsOH is sometimes used in place of NaOH for homodyne SHG measurements. The solution pH is measured for each salt concentration and shown to remain largely constant (i.e., only differing by ± 0.05) over the range of ionic strengths investigated here. For HD-SHG measurements, pure Millipore water (18.2 MΩ*cm) and alkali halide solutions are equilibrated with $CO_2$ overnight to achieve a stable pH of 5.8 throughout the experiments. We obtained NaI from Santa Cruz Biotechnology (Lot # J1515, Catalog # sc-203388A, ≥99% pure), NaCl from Sigma-Aldrich (Part # 746398-2.5KG, Lot #

SLBK2618V, ≥99% pure) and Alfa Aesar (Lot # M08A016, ≥99% pure), NaBr from Sigma-Aldrich (Part # 310506-100G, Lot # MKBQ8200V, ≥99% pure), RbCl from Aldrich (Part # R2252-50G, Lot # WXBC0662V, ≥99% pure), KCl from Sigma-Aldrich (Part # 746435-500G, Lot # SLBP3785V, ≥99% pure), CsCl from Sigma-Aldrich (Part #s C3011-25G, C3011-100G, Lot # SLBP4992V and ≥99% pure) and Aldrich (Part # 203025-10G, ≥99.999% pure), LiCl from Dot Scientific (Lot # 46084-64653, >99% pure), CsI from Aldrich (Part # 202134-25G, Lot # MKBX2413V, ≥99.9% trace metal basis), CsOH from Aldrich (Part # 516988-25G, Lot # MKBX2444V and 99.95% trace metal basis), NaOH from EMD Chemicals (Lot # B0312669 941) and HCl from Fisher Scientific (Lot # 155599).

**III. Results and Discussion.**

**A. Ion Specific Outcomes in Homodyned SHG $\chi^{(3)}$ Measurements.** We first present and discuss the influence of alkali and halide ions on the SHG response from the silica/water interface at circumneutral pH. Our earlier studies using NaCl[65] revealed two response regimes for the homodyne-detected SHG signal change, specifically, an initial increase followed by a gradual decrease when screening from low to high ionic strength at the silica/water interface. The increase in response Regime I is observed for ionic strengths below about 0.1 mM to 1 mM and results from interference effects of the changing $\chi^{(3)}$ phase angle and its coefficient.[13, 49-52, 57] Fig. 1A and Fig. 1B show that the alkali halides studied here display the same two response regimes as those previously reported for NaCl. In response Regime I, the detected SHG responses increase by up to 15% when compared to the condition of zero salt added for all salts studied except for CsCl (vide infra). Response Regime II occurs for ionic strengths >1 mM. Here, use of any combination of cations and anions produces SHG response decreases by as much as 50% when compared to the condition of zero salt added, following well-characterized charge-screening behavior.[13, 50, 66] The extent of the SHG response decay with increasing salt concentration appears to trend with cation and anion softness. Fitting eqn. 2 to the SHG intensities recorded at the PMT after correcting them for $\chi^{(3)}$ interference effects (Fig. 2) reveals that the interfacial charge density point estimates range between -0.0039 (NaI) and -0.016 C/m² (NaCl) for the sodium halides surveyed when we consider one standard deviation (Table 1). The charge density point estimates obtained from the fits of the chloride salt data range from -0.0017 (CsCl) to -0.016 C/m² (NaCl). Varying the value of A used in equation 2 to correct for the $\chi^{(3)}$ interference by 5% in a sensitivity analysis we carried out did not influence the trend seen for the surface charge densities of the alkali halides studied (Tables S1 and S2). The charge densities obtained for the fused silica/water interface at pH 7 we report here using SHG support this conclusion (Table 1), albeit for considerably lower sheer rates than what is possible in a liquid jet: the magnitude of the charge density at the fused silica/water interface decreases by a factor of ~3 when we replace NaCl with NaI at the interface. On the other hand, the magnitude of the charge density drops by a factor of ~5 when comparing NaCl to RbCl. Liquid-jet XPS experiments by Brown and co-workers indicate that the negatively charged silica surface interacts more with cations than anions.[14, 67] The surface potential was reported to increases by a factor of ~1.5 (from ~ -260 mV to ~ -400 mV) when the electrolyte at the interface was switched from CsCl to NaCl at 50 mM ionic strength at pH 10,[15] while their follow up study[68] demonstrated anion identity has no significant influence on the surface charge density of silica nanoparticles. Though XPS studies by Brown et al. indicate cation identity influences surface potential to a considerable extent, SFG studies by Yang et al. showed that the structure of water at the fused silica/water interface at an ionic strength of 50 mM is not affected by change in alkali ($Li^+$, $Na^+$, $K^+$) ion with chloride as the anion at pH ~5.8.[29] Those results are in agreement with molecular dynamics simulations demonstrating that monovalent cation ($K^+$, $Na^+$) and anion ($Cl^-$, $I^-$) effects on water molecules orientation to the fused silica interface are due to inner and outer sphere complexes formed, respectively.[69] Another computational study has shown that hydrated $Na^+$ and $Cs^+$ ions adsorb to a negatively charged silica surface via inner and outer sphere mechanisms, respectively.[70] Yet another study by Franks et al.[71] reported that $Cs^+$ and $K^+$ hydrated ions adsorb in greater quantity at the silica/water interface than $Li^+$ and $Na^+$ hydrated ions; thus the $Cs^+$ ions produce a less negative zeta potential at the silica/water interface. Other electrophoretic measurements have confirmed that the magnitude of the negative zeta potentials at the silica/water interface changes in accordance with the Hofmeister series ( $Cs^+ < K^+ < Na^+ < Li^+$).[72-76] A SFG study of the silica/water interface by Hua et al. reported that $Li^+$ and $NH_4^+$ perturb the hydrogen bonding network of water whereas $Na^+$ and $K^+$ have little effect on the water structure at the silica/water interface. That study showed the intensity of the SFG signal at the silica/water interface trends with $Li^+ \sim Na^+ > NH_4^+ > K^+$.[77]

At a specific relative permittivity, interference corrections and the GC model depend solely on ionic strength; therefore, any observed differences between salts indicate changes of either the system's $\chi^{(3)}$ or $\chi^{(2)}$ values, i.e. chemistry in the electrical double layer. Indeed, when an aqueous salt solution is present at the charged silica/water interface, we expect the hydrated cations in solution to adsorb directly to the negatively charged fused silica surface within the electrical double Stern layer. Possible structures have been inferred recently by DFT calculations by Gaigeot and co-workers, who reported that $Na^+$ forms an inner sphere complex with four water molecules at the fused silica/water interface.[59] Another computational study has demonstrated that hydrated $Na^+$ and $Cs^+$ ions adsorb to a negatively charged silica surface via inner and outer sphere mechanisms respectively.[70] We therefore propose that the adsorption of the different hydrated cations within the electrical double Stern layer causes varying changes in the $\chi^{(3)} \Phi_0 / \chi^{(2)}$ ratio which in turn influences the level of SHG signal rise observed due to phase matching when the aqueous solution is held at pH 7 and low ionic strength.

A recent modeling study[78] relied on the accurate knowledge of the $\chi^{(3)} / \chi^{(2)}$ (i.e. B/A) ratio, which they termed as s. Using a survey of the literature, they estimated an s value of ~120 $V^{-1}$ for the air/water interface. For the buried interfaces of interest here, strong ordering at the interface would produce a larger $\chi^{(2)}$, decreasing the value of *s*. Indeed, one recent report[79] of the silica/water interface in the presence of NaCl found $s$ = -17.5 V$^{-1}$ at an unspecified pH, presumably 5.8 (water equilibrated with laboratory air). Table 1 shows similar absolute values, at lest within an order of magnitude and for conditions of pH 7 and low (0.2 mM) ionic strength. The variance of these parameters with different salts implies different structuring directly at the interface possibly characterized by $\chi^{(3)}/\chi^{(2)}$, as recent theoretical studies have suggested.[7, 53, 80] Using the charge densities (Table 1) obtained from fitting the data in Fig. 1 with varying relative permittivity values from 1 to 80, we then calculated the $\chi^{(3)} \Phi_0/\chi^{(2)}$ ratios at 0.2 mM (200 µM) for the alkali halides studied here at pH 7 (Fig. 3). While the $\chi^{(3)} \Phi_0/\chi^{(2)}$ ratios were found to be ~1 for the six alkali halides, we find statistically significant differences for some of the cations and anions for all relative permittivity values surveyed here.

To further explore ion-specificity of the SHG $\chi^{(3)}$ process for alkali halides, we conducted on/off SHG experiments with NaCl, RbCl, LiCl, and NaI at an ionic strength of 0.2 mM and pH 7 showing that cations (Fig. S1A, C) dictate the rise of the SHG increase – owing to phase matching at low ionic strength – more than anions (Fig. S1B, C). We note that the magnitudes of the SHG increases at low ionic strength trend with the hardness of cation present (Fig. S1), like what was observed in Regime I of the equilibrium adsorption isotherms (Fig. 1). These experiments are consistent with the observation that the SHG intensity in this regime is reversible.[13] Yet, we note that an exception is observed for CsCl, for which half of the trials carried out produced either the opposite or no change in SHG signal intensity in this salt concentration regime (see Fig. S2).

**B. Heterodyne-detected SHG $\chi^{(3)}$ Measurements Separate the $\chi^{(3)} \Phi_0$ and $\chi^{(2)}$ Contributions for NaCl, RbCl, and NaI.** Our recent HD-SHG report demonstrates how to separate the $\chi^{(3)} \Phi_0$ and $\chi^{(2)}$ contributions to the SHG signal at constant pH and varying ionic strengths.[42] Now, we use this method to determine how the cations or anions studied here influence the $\chi^{(3)} \Phi_0$ and $\chi^{(2)}$ contributions to the SHG signal at low ionic strength (0.2 mM) and pH 7. Following the same approach reported earlier on fitting and optimizing interference scans (Fig. 4A-C) for NaCl at varying ionic strengths, we derived $E_{sig}$ and $\Delta\varphi_{sig}$ ($\Delta\varphi_{sig} = \varphi_{sig,0.2\,mM} - \varphi_{sig,100\,mM}$) for 0.2 mM NaCl, 0.2 mM RbCl, and 0.2 mM NaI (see Table 2) by fitting the recorded interference patterns to cosine functions as described earlier.[42] The amplitude of the fitted cosine function for each interference scan is directly proportional to $E_{sig}$, and changes in the phase of the fit cosine function can be attributed to changes in $\varphi_{sig}$ for each ionic strength.

In our analysis, we assume that the signal phase angle at 100 mM, $\varphi_{sig,100\,mM}$, is 0° since $\chi^{(3)}$ and $\chi^{(2)}$ are purely real and the interfacial potential is minimized to a few mV. Moreover, eqn. 4 shows that the DC phase angle, $\varphi_{DC}$, is almost zero at 100 mM ionic strength.[45] Our previous work found a reversible phase shift ($\Delta\varphi_{sig}$) of 19.1° ± 0.4° between 100 mM NaCl and deionized water ($CO_2$ equilibrated and at a measured pH of 5.8).[42] We note that this value was obtained from experiments that initially cycled between deionized water and 100 mM NaCl solution at least three times before performing NaCl screening (going from low to high ionic strength) so as to overcome hysteresis effects[41, 81] resulting from screening from low to high ionic strength. In contrast, our current method does not expose the hemispheres to high salt before cycling between pH 5.8 water and 0.2 mM salt. Data collected for several runs on our phase-referenced system shows that there is a consistent phase shift of 29.7 ± 0.7° (see Fig. S3) between 100 mM NaCl and $CO_2$ equilibrated water at pH 5.8 for the first jump (from water to high salt) when working with a freshly prepared fused silica hemisphere. Therefore, we set the phase of pH 5.8 water at 29.7° ± 0.7° to then find $\Delta\varphi_{sig}$ for 0.2 mM of each salt. We further assume that $\varphi_{sig,100\,mM}$ is the same for all the salts studied. Indeed, we conducted duplicate flow experiments (see Fig. S4) on our phase-referenced system where we cycled between 100 mM NaI, 100 mM NaCl, and 100 mM RbCl. By fitting the resulting interference scans, we found $\Delta\varphi_{sig} \leq 2°$ for any pair of the salts (NaCl, RbCl, and NaI) at 100 mM. Thus, our assumption is valid within the current measurement uncertainty of our HD-SHG spectrometer. The trend of the $E_{sig}$ values obtained here generally agrees with the SHG signal trend seen for the on/off traces performed for 0.2 mM alkali ions present at the fused silica/water interface (see Fig S1) using our homodyne detection system.

The values in Table 2 and $\varphi_{DC}=13.3°$ (again, computed using eqn. 4 for a Debye length at 0.2 mM ionic strength for all salts used here) were then put into equations 5 and 6 to obtain $\chi^{(3)}\Phi_0$ and $\chi^{(2)}$. Fig. 4 shows similar $\chi^{(3)}\Phi_0$ values for RbCl, NaCl, and NaI at 0.2 mM while the $\chi^{(2)}$ values we calculated were found to increase in the order of NaI > NaCl > RbCl (the $\chi^{(2)}$ values are related as follows: $\chi^{(2)}_{RbCl} = \frac{1}{2}\chi^{(2)}_{NaCl} = \frac{1}{4}\chi^{(2)}_{NaI}$). The corresponding Argand diagrams are shown in Fig. 5. These results support, for the first time through a direct measurement, the aforementioned notion that the $\chi^{(2)}$ term originates mainly from the Stern layer whereas the $\chi^{(3)}\Phi_0$ term originates from the diffuse layer. Our finding that $\chi^{(3)}\Phi_0$ is invariant with the type of alkali halide present at these low ionic strengths is consistent with recent work by Pezzotti et al., who calculated the SFG $\chi^{(3)}$ spectra for differing concentrations of KCl, Cl[-], and I[-] ions, ranging between 0.4M and 2M, at the air/water and quartz/water interfaces.[53] The resulting spectra were independent of ion type, concentration, or interface.[53] Our results are also fully consistent with $\chi^{(3)}$ responses reported by Wen et al. from HD-SFG measurements of the air/water interface that are invariant with a variety of surfactants and pH conditions.[42]

**IV. Conclusions.** We have probed the charged fused silica/water interface at pH 7 using homo- and heterodyne detected SHG $\chi^{(3)}$ spectroscopy for ionic strengths ranging from 10 μM to several 100 mM for NaCl, NaBr, NaI, KCl, RbCl, and CsCl. Below about 1 mM, SHG signal intensities were observed to increase by up to 15% when compared to the condition of zero added salt due to phase matching for NaCl, NaBr, NaI, KCl, and RbCl. The magnitude of the SHG increase was found to trend with the hardness of the alkali ions present. For ionic strengths above 1 mM, use of any combination of cations and anions was found to coincide with SHG E-field decrease by as much as 50% when compared to the condition of zero salt added. The decrease in SHG response followed a well-characterized charge screening behavior that was found to trend with ion softness. When the obtained SHG intensities were corrected for phase matching and then fitted using the Gouy-Chapman model, the magnitude of the charge density at the fused silica/water interface was found to decrease by a more substantial factor when changing the electrolytic alkali ion from $Na^+$ to $Cs^+$ than when changing the electrolytic halide ion from $Cl^-$ to $I^-$. Data obtained from phase-referenced SHG measurements indicate the diffuse layer is independent of ions present, but Stern layer properties, as reported by $\chi^{(2)}$, are clearly subject to ion specificity for the ions surveyed in this work.

Despite these advances, it is not known how charge shielding through, for instance, ion pairing in crowded interfacial environments impacts the relative permittivity, $\varepsilon_r$, there. Theoretical and computational analyses show $\varepsilon_r \ll 80$ at least for distances close (<10 nm) to the interface,[82-88] recapitulating experiments reported for flat[13] and nano-confined[89] aqueous interfaces. Ion specificity is likely to be important for this parameter as well. An outstanding question concern the electrolyte valency, $z_i$, as a reasonable descriptor for conditions where ion correlations, such as pairing, crowding, or charge shielding, are expected to be important. Moreover, we are interested in parameterizing ion correlations, i.e. ion pairing, contact ion pair formation, from the experimental data obtained using the new HD-SHG spectroscopic method presented here.

**Supplementary Information.** Derivation of the equations used in this work; normalization procedure for homodyne SHG intensities; additional Gouy-Chapman fit results; homodyne results for CsCl; and $\varphi_{sig}$ values obtained for 0.1 M RbCl, 0.1 M NaCl, and 0.1 M NaI.

**Acknowledgments.** MDB gratefully acknowledges support from the PPG fellowship program at Northwestern University. This work was supported by the US National Science Foundation (NSF) under its graduate fellowship research program (GRFP) award to PEO. PEO also gratefully acknowledges a Northwestern University Presidential Fellowship. FMG gratefully acknowledges NSF award number CHE-1464916.

**Author Contributions.** MDB, PEO, EL, and JV performed the experiments. MDB, PEO, JV, EL, and FMG analyzed the data. The manuscript was written with substantial contributions from all authors.

**Author Information.** The authors declare no competing financial interests. Correspondence should be addressed to FMG (geigerf@chem.northwestern.edu).

**Tables and Table Captions**

**Table 1.** Fit parameters ($\chi^{(3)}/\chi^{(2)}$ or B/A and $\sigma$) obtained from fitting the Gouy-Chapman model (eq. (S3)) to the charge screening isotherms for all six alkali halides surveyed here. Values in parentheses indicate the 1-sigma uncertainty associated with the point estimate. Average SHG E-field values (A) for ionic strengths ≥0.1M are used for phase matching correction for the respective alkali halides.

| Alkali Halide | $\chi^{(3)}/\chi^{(2)}$ (V$^{-1}$) | $\sigma$ (C/m²) |
|---|---|---|
| NaCl | -4.1(3) | -0.014(2) |
| NaBr | -10(2) | -0.0075(13) |
| NaI | -10(2) | -0.0045(6) |
| KCl | -12(2) | -0.0052(7) |
| RbCl | -14(1) | -0.0030(2) |
| CsCl | -20(2) | -0.0019(2) |

**Table 2.** Values of $\Delta\varphi_{sig}[°]$ and $E_{sig}$ obtained from fitting interference patterns obtained from phase-referenced measurements of 0.2 mM alkali halide solutions in contact with fused/water interfaces at pH 5.8. $E_{sig}$ values reported are referenced to the $CO_2$-equilibrated water response. $\Delta\varphi_{sig}[°]$ values are calculated with $\varphi_{sig, H_2O,\ pH\ 5.8} = 29.7° \pm 0.7°$ assuming that $\varphi_{sig, 100\ mM} = 0°$. Further details about this assumption are discussed in the text.

|  | Alkali halide | | |
| --- | --- | --- | --- |
| **Parameter** | **RbCl** | **NaCl** | **NaI** |
| $\Delta\varphi_{sig}[°]$ | 12.0±0.8 | 11.1±0.8 | 9.3±0.8 |
| $E_{sig}$ | 1.18±0.03 | 1.36±0.02 | 1.43±0.03 |

**Figures and Figure Captions**

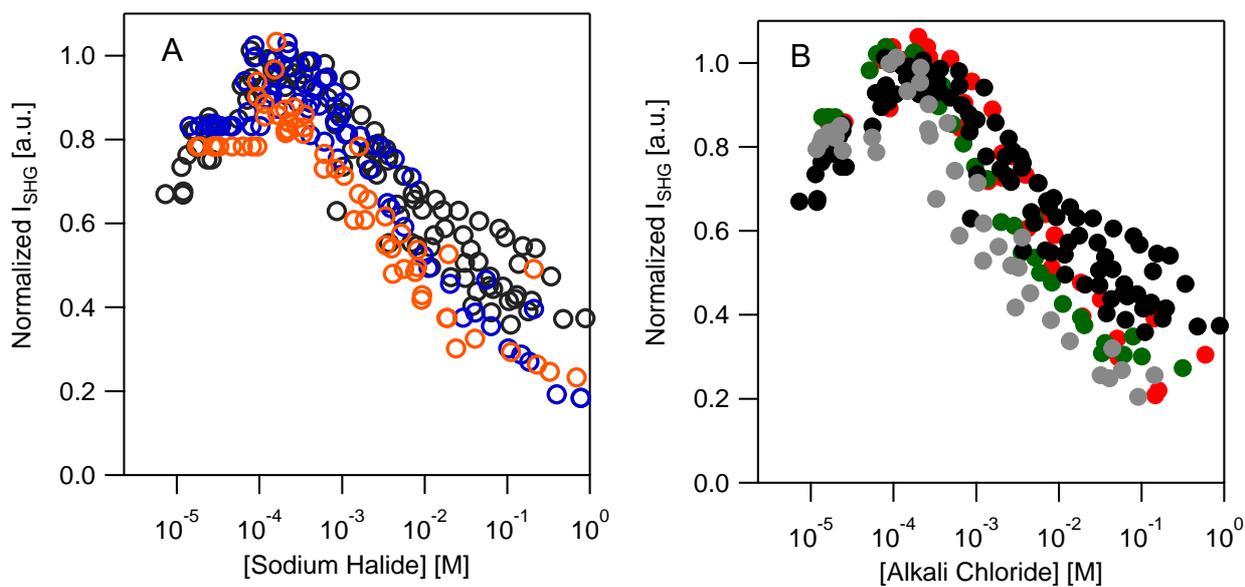

**Figure 1.** Homodyne-detected SHG intensity as a function of ionic strength for (**A**) anion series (NaCl, black, NaBr, blue, and NaI, gold) and (**B**) cation series (NaCl, black, KCl, red, RbCl, green, and CsCl, gray) at the fused silica/water interface. All data was collected at pH 7 and at a flow rate of ~1 mL/s. Please see text for details.

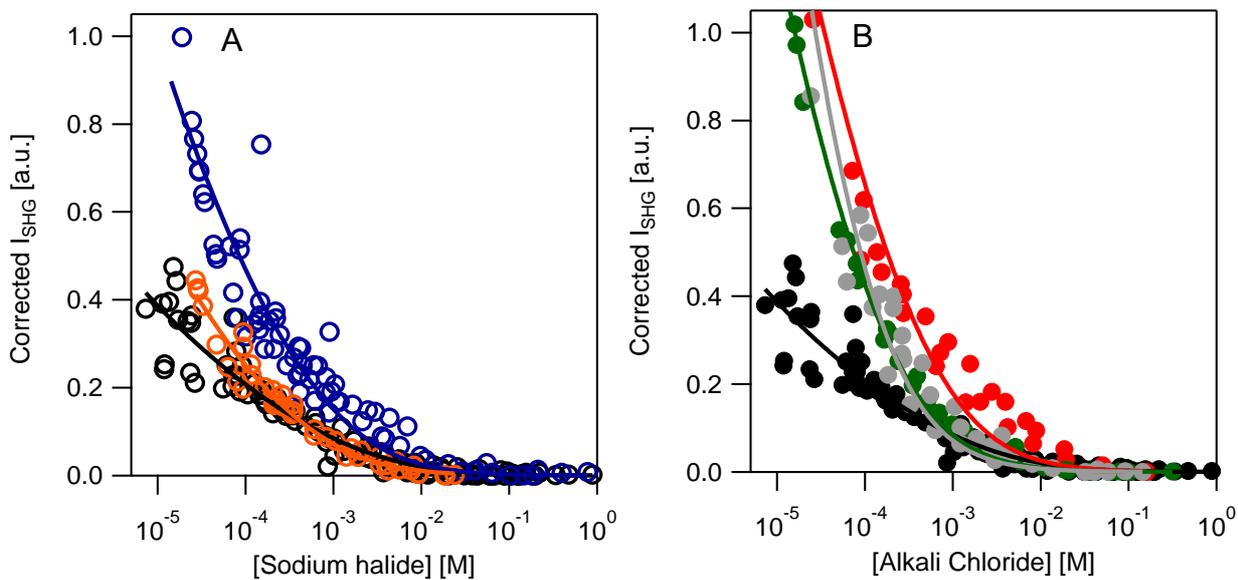

**Figure 2.** Gouy-Chapman model fits (solid line curves) for corrected SHG intensities from Fig. 1 (circles) versus ionic strengths for **(A)** anion series (NaCl, black, NaBr, blue, and NaI, gold) and **(B)** cation series (NaCl, black, KCl, red, RbCl, green, and CsCl, gray) at the fused silica/water interface. Please see text for details.

Boamah et al.Page 25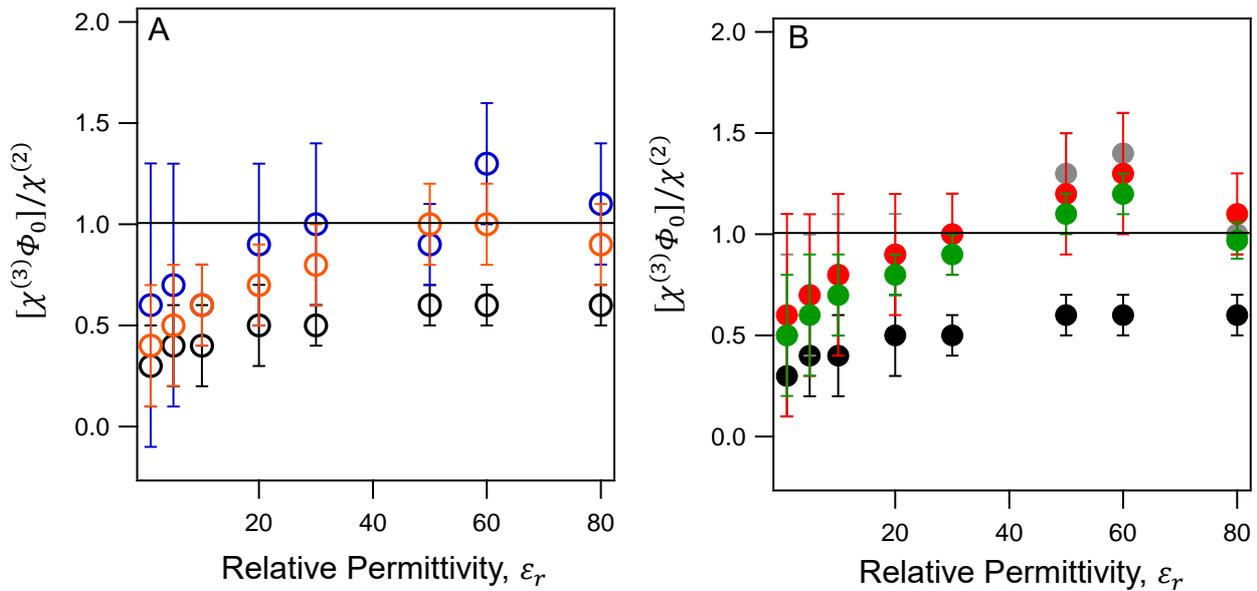

**Figure 3.** Ratio of $\chi^{(3)}\Phi_0/\chi^{(2)}$ calculated from the experimental homodyne-detected SHG data from Fig. 1 for sub mM ionic strength at varying relative permittivity for (**A**) sodium halides (NaCl – black/bottom, NaI – gold/middle, and NaBr – blue/top) and (**B**) alkali chlorides (NaCl – black/bottom, RbCl – green/second-lowest, KCl – red/third highest, and CsCl – gray/top). $\Phi_0$ values are calculated using an ionic strength of 0.2mM and the charge densities obtained from the fits in Fig. 2 at varying relative permittivities. Error bars represent the propagated uncertainties from the standard deviations. Horizontal black lines represent $\chi^{(3)}\Phi_0/\chi^{(2)} = 1$. Here, $\chi^{(3)}\Phi_0/\chi^{(2)}$ values have arbitrary units.

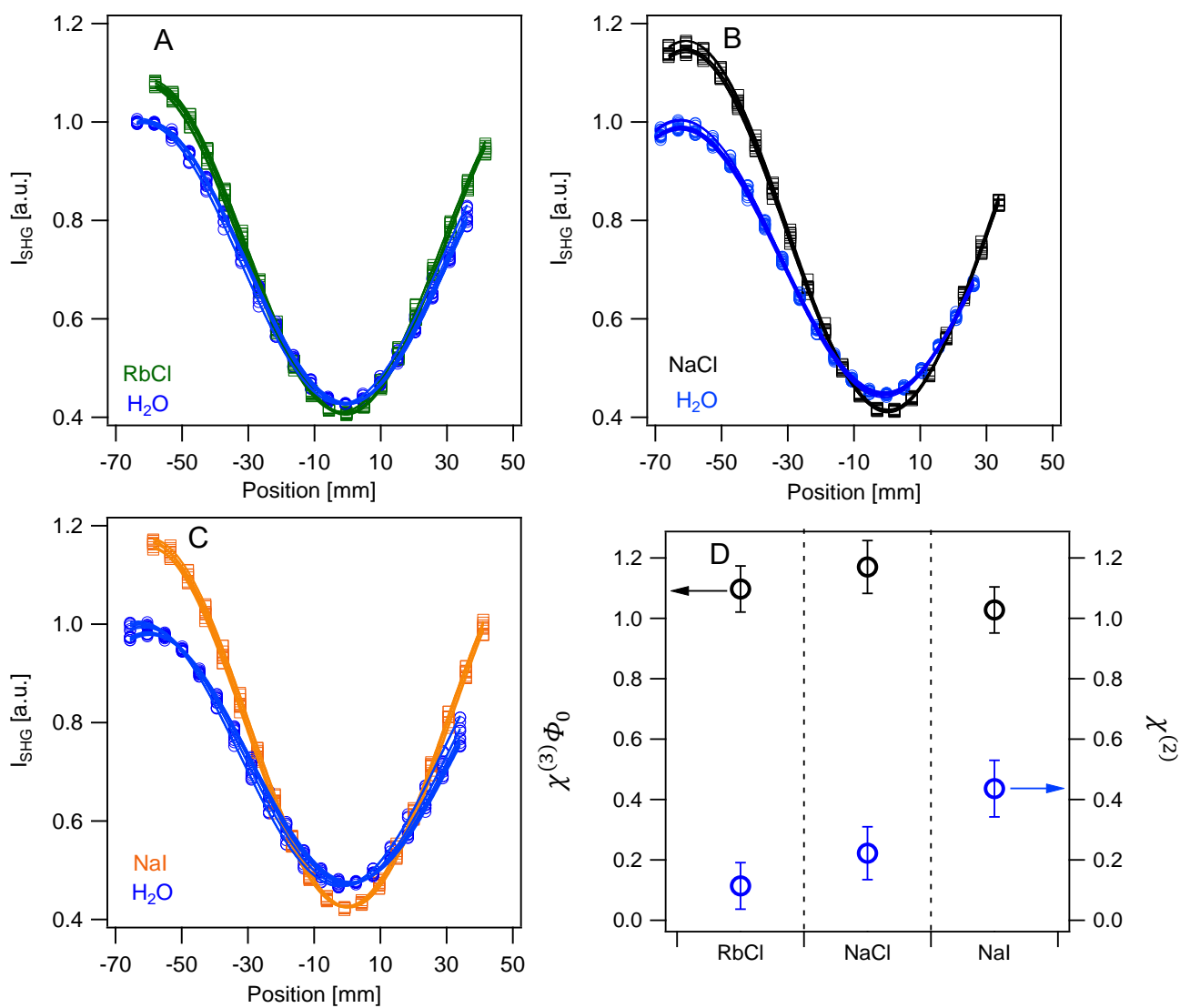

**Figure 4**. Triplicate measurements of interference fringes recorded for fused silica/water interfaces in contact with $CO_2$-equilibrated water (empty blue circles) and 0.2 mM **(A)** RbCl, **(B)** NaCl, and **(C)** NaI. Average $\Delta\varphi_{sig}$ [°] and $E_{sig}$ values obtained from the fringes are reported in Table 2. Distances are that of the local oscillator and the signal source at the fused silica/water interface. **(D)** Graph showing the separated $\chi^{(3)} \Phi_0$ (left axis, black circles) and $\chi^{(2)}$ (right axis, blue circles) contributions for RbCl, NaCl, and NaI at an ionic strength of 0.2mM.

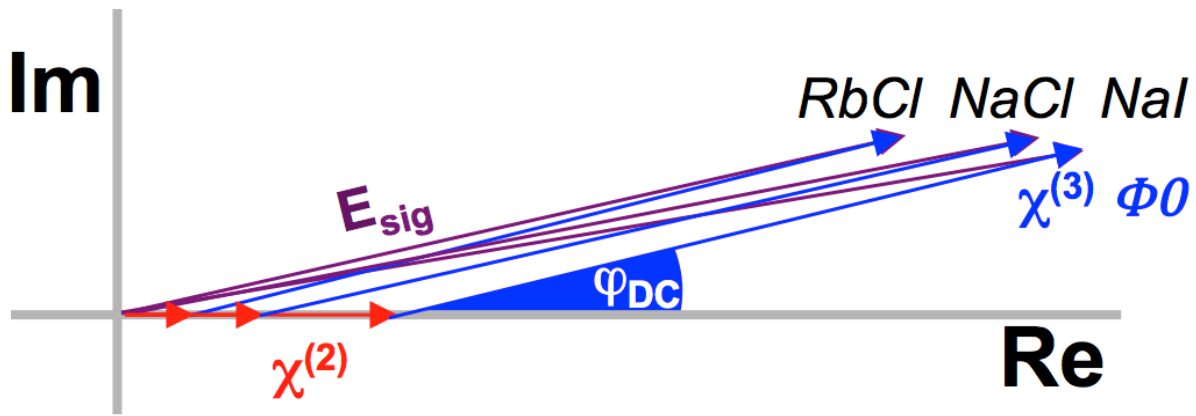

**Figure 5**. Argand diagram for ion specific SHG responses for 0.2 mM RbCl, NaCl, and NaI solutions maintained at circumneutral pH interacting with the fused silica/water interface. At constant ionic strength, $\varphi_{dc}$ is constant. The variations in the measured values of $E_{sig}$ and $\varphi_{sig}$ results in a largely constant magnitude of $\chi^{(3)}\Phi(0)$ while that of $\chi^{(2)}$ varies for the different salts.

TOC Graphic

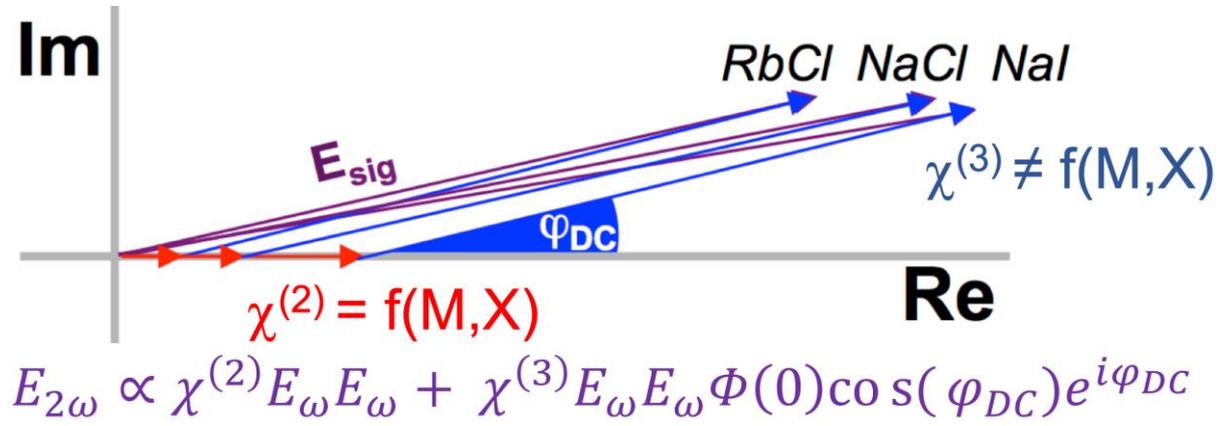